\documentclass[mathleft]{an}  

\usepackage{graphicx}
\usepackage{times}
\usepackage{natbib}
\bibpunct{(}{)}{;}{a}{}{,} 
\def\msun{M$_\odot$}

\def\hu{HU\,Aqr}

\begin{document}

\Pagespan{000}{}
\Yearpublication{2014}%
\Yearsubmission{2013}%
\Month{12}%
\Volume{000}%
\Issue{00}%

\title{On the ephemeris of the eclipsing polar HU Aquarii}

\author{A.D.~Schwope\inst{1}\fnmsep\thanks{Corresponding author:
  \email{aschwope@aip.de}\newline}
\and
B.D.~Thinius\inst{2}
}
\titlerunning{On the ephemeris of HU Aqr}
\authorrunning{A.D.~Schwope \& B.D.~Thinius}
\institute{Leibniz-Institut f\"ur Astrophysik Potsdam (AIP),
              An der Sternwarte 16, 14482 Potsdam, Germany
              \and
Inastars Observatory, Hermann-Struve-Str. 10, 14469 Potsdam, Germany
}

\received{}
\accepted{}
\publonline{}

\abstract{The magnetic cataclysmic variable HU Aquarii displayed pronounced quasi-periodic 
modulations of its eclipse timing.
These were interpreted in terms of the light-travel time (LTT) effect caused by a 
circumbinary planet or planetary system. 
We report new photometric observations that revealed another precise eclipse timing for the 
October 2013 epoch, the first obtained in a high accretion state after many years 
in low or intermediate states. 
The eclipse was observed to occur earlier by $95.3\pm2.0$\,s or $62.8\pm2.0$\,s than expected
for an assumed linear or quadratic ephemeris, respectively. The implied
apparent strong evolution of the orbital period calls for a revision 
of the current planetary model or the planetary parameters. The object deserves 
further monitoring to uncover the true nature of the observed variability 
and to constrain the properties of the proposed planet or planetary system.
The new observations prove that advanced amateur equipment can successfully be used in the
growing field of planet search in wide circumbinary orbits via the LTT effect.
}


\maketitle
%

\section{Introduction}
\hu\ is an eclipsing magnetic cataclysmic variable with a 125.0\,min orbital period. 
When discovered in 1993 as the optical counterpart to the soft X-ray and EUV sources RX\,J2107.9-0518/RE2107-05
\citep{hakala+93, schwope+93} it was the brightest eclipsing object displaying 
the most extended eclipse. Those properties triggered broad observational studies
to disentangle accretion phenomena and the accretion geometry in a strongly 
magnetic environment. Particular emphasis was given to model the detailed eclipse structure. 

Comprehensive X-ray and EUV observations with the ROSAT and EUVE 
satellites took place  between 1992 and 1998 \citep{schwope+01}. 
These studies established the eclipse egress as fiducial mark to determine
the orbital period and a long-term ephemeris. The ingress into eclipse was found to be strongly affected 
by absorbing matter in the accretion curtain preventing an unequivocal 
determination of eclipse centre. 
The 31 epochs given for the eclipse egress that were based on soft X-ray observations 
already displayed systematic offsets with respect to a linear 
ephemeris of $\pm5$\,s. The size of the observed effect was still compatible with a migration of the
accretion spot over the surface of the white dwarf. 

Monitoring observations with high-time resolution were continued at 
X-ray and optical wavelengths in the new millennium. \cite{schwarz+09} presented a further set of 62 
eclipse epochs and were the first to discuss the timing residuals, that were then 
larger than the size of the white dwarf,  in terms of an unseen third body
and derived a possible mass of $M_3 \ga 5 M_{\rm Jup}$ for a planetary companion.

\cite{qian+11} added a further 11 eclipse epochs and claimed the discovery of a 
circumbinary planetary system around the accreting binary. This hypothesis is 
debated. Their data seem to be offset from a more comprehensive data set 
presented by \cite{godz+12} and the proposed planetary system seems to 
loose stability on relatively short time scales \citep{horner+11,wittenmyer+12}. The
study by \cite{hinse+12} on the other hand supports at least a two-planet scenario. 
Opportunities to reconstruct the evolution of the close binary through the presence 
of a planetary system were studied by \cite{port-zwart13}.
The stellar and binary parameters were updated by \cite{schwope+11} who derive 
a total mass of the close binary of $0.98 \pm 0.10$\,\msun. 

The frequent low- and intermediate accretion states of the object prevented 
further detailed X-ray observations during the past 12 years. 
In July 2013 a high accretion 
state was noticed by monitoring observations with the robotic telescope STELLA 
at Tenerife. Swift and XMM-Newton observations were triggered to study the 
X-ray emission in a high accretion state. The observations with XMM-Newton were 
scheduled immediately after re-appearance of the source
in the visibility zone of XMM-Newton and took place eventually on October 25. 

At that occasion attempts were made to obtain simultaneous ground-based data 
(Schwope et al.~2014, in preparation). 
Here we report quasi-simultaneous optical observations that were obtained 
utilizing the private equipment of the Inastars observatory Potsdam 
(IOP, B15, N 52.42392, E 13.012892).


\section{Observations and data reduction}
\hu\ was observed with a 14 inch Celestron reflector of Schmidt-Cassegrain type 
during six clear nights in October 2013.
The telescope is permanently installed on the roof of a one-family dwelling
in a suburb of Potsdam, the capital of the state of Brandenburg (Germany).    
An Optec NextGEN WideField 0.5X telecompressor revealed an aperture
ratio of f/5.5 and a field of view of $24\times16$\,minutes of arc.
The data were obtained through an 
ASTRONOMIK\footnote{http://www.astronomik.com/de/photographic-filters/cls-ccd-filter.html} 
filter which blocks typical emission lines at light-polluted sites 
and has almost full transparency in two windows 
that range from $450-520$\,nm and $640-690$\,nm, respectively.  
The camera used, an SBIG ST-8XME, features  
a CCD with $1530\times 1020$ pixels (pixel size $9\times9$ microns)
and was used mostly with a $2\times2$ binning.
The temperature of the CCD is actively controlled and was set to $-25.0$\degr C
during the observations which was achieved with an rms deviation 
of about 0.3\degr C.  
The telescope/camera system is mainly used for observations of minor planets.
Several new discoveries are confirmed by the minor planet center (MPC) of the IAU
(see e.g.~http://thinius.net/discstatus.htm.)
Since this was the first time, that the equipment was used for 
time-resolved photometry with an as high as possible resolution, some experiments 
were made using different CCD binning
and exposure time per frame to find a compromise between highest possible
time resolution and an acceptable signal-to-noise ratio. 

A summary of the observations reported in this paper is given in Table~\ref{t:log}, which 
lists the exposure time, the achieved time resolution, the number of frames 
and the time interval covered during each of the nights.
During the last night, beginning October 30, a total of 1995 science frames were obtained.
Almost all of those, 1973 frames, were taken with an exposure 
time of 4\,s. During the two eclipses, however, the exposure time was 
changed to 20\,s (7 frames) and 60\,s (15 frames) in an attempt to more securely
measure the remaining brightness of the binary. Those modifications 
of the observing sequence 
resulted in a corresponding less accurate determination of the eclipse time 
(see Tab.~\ref{t:ecl}). 

The start time of each observation 
and the exposure length was written into the fits headers generated upon read-out.
In the context of the current paper the timing accuracy of the data is crucial.
The computer equipment was correlated with a time signal of atomic clocks  
every five minutes via the Network Time Protocol (NTP). The synchronization process
runs on the measuring computer with high priority. For time synchronization  
times servers in Germany, the USA as well as India were used. The list of 
potential time servers comprises 10 redundant stations. The time difference
between time server and PC was logged for each synchronization run. Different 
package transmission times were reported for the servers located at different continents, 
the measured time differences between server and PC however were found to be 
stable with a scatter of about 10 ms or even better.

The CCD software writes images with 1 second time resolution by cutting the fraction 
of a second. The times written to the CCD headers are thus biased by $-0.5\pm0.5$\,s.
Shutter latency measurements gave a mean latency of $0.77\pm0.02$\,s. 
Hence, all times measured on the original data were corrected afterwards by $+1.27$\,s
and a systematic uncertainty of 0.52\,s was added to the estimated statistical 
uncertainties. Confidence that the systematic uncertainties are not larger 
than this value is derived from the positions of minor planets that were 
determined with the same instrumental setup. The positions reported to 
the IAU-MPC agree with the orbit predictions always better than one arc-second.
  
The raw CCD frames were bias-corrected, dark-subtracted and flat-fielded using 
calibration data obtained during each of the nights. For this purpose 
average frames of each 50 sky-flats, bias- and dark-frames taken immediately 
before the observations with the same instrumental setup and CCD 
temperature were used. No trend in the bias values was recognized.

Afterwards, differential photometry using ESO-MIDAS routines was performed
with respect to comparison star 'C'  
 \cite[USNO B1.0 0846-0616384,  2MASS coordinates RA(J2000) = 21:07:59.091, 
DE(J2000) = $-$05:18:49.62, 14.696mag, ][]{schwope+93}.
Firstly, the centroids of the stellar images of both the target and the comparison star
were determined via two-dimensional 
Gaussian fits on a number of images with good detections of both stars to determine 
the pixel offset between them.
Then aperture photometry was performed on each CCD frame using concentric apertures for 
object, nomansland, and sky, centered initially on the comparison star, then at the 
offset position using the same apertures.
With this procedure a brightness measurement was 
possible even during eclipses of HU Aqr, when the source was too faint 
for centering. Errors of individual measurements
were determined taken into account the sky brightness, the object brightness 
and the read-noise properties of the CCD. The quantity further investigated
is the flux ratio between the target and the comparison star, also referred to as 
relative flux.

\begin{table}[t]
\caption{Time-resolved photometric observations of \hu\ obtained at Inastars Observatory 
in October 2013. Given are the observation date, time interval covered, exposure time, 
time resolution, and the number of frames obtained per night.
\label{t:log}}
\begin{tabular}{lcccr}
\hline
Date & Interval & $t_{\rm exp}$ & $\Delta t$ & \# frames \\
& JD +2456588. & [s] & [s] \\
\hline
20131022 &  $0.29217 - 0.36269$ & 3.0 & 6.1 & 976 \\
20131023 & $ 1.31179 - 1.37931$ & 3.0 & 6.1 & 277 \\
20131024 & $ 2.24323 - 2.42821$ & 4.0 & 6.9 & 1903 \\
20131026 & $ 4.26040 - 4.38630$ & 8.0 & 10.0 & 677 \\
20131029 & $ 7.27689 - 7.39935$ & 2.0 & 3.8 & 1951 \\
20131030 & $ 8.23877 - 8.41131$ & 4.0 & 6.9 & 1995 \\
\hline
\end{tabular}
\end{table}

\begin{figure}[t]
\resizebox{\hsize}{!}{\includegraphics[clip=]{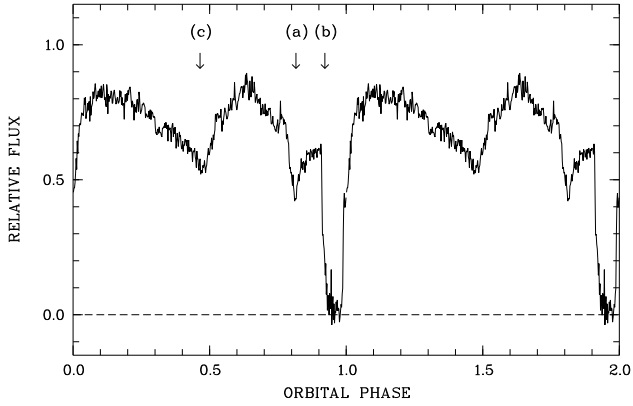}}
\caption{Phase-averaged light curve of \hu\ obtained October 2013
at Inastars Observatory (500 phase bins). Orbital phase refers to a linear
ephemeris ${\rm BJD(TDB)} = 2449102.92004 + E \times 0.086820400$
}
\label{f:lcfol}
\end{figure}

\begin{figure}[t]
\resizebox{\hsize}{!}{\includegraphics[clip=]{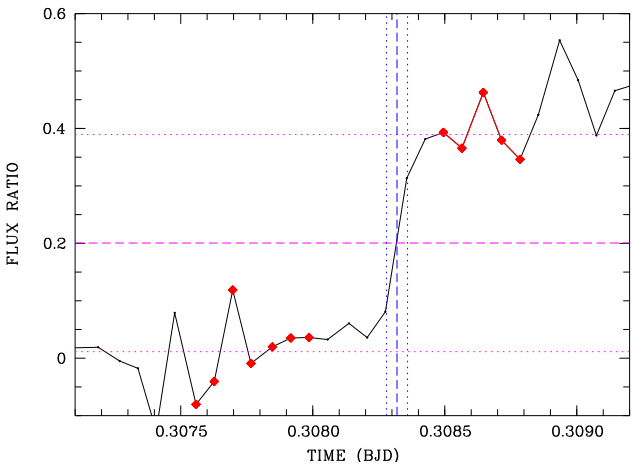}}
\caption{Determination of eclipse egress on October 24. One data point
represents a 4\,s integration, the time resolution is  
6.9 s. Shown with dotted vertical lines are mean eclipse and out-of-eclipse
values used to compute the half-light value using data shown with larger 
symbols. The measured time of egress was read at half light and shown with 
a vertical dashed line. The timing error was set to half a resolution element
and is indicated by vertical dotted lines.}
\label{f:meas}
\end{figure}

\begin{figure}[t]
\resizebox{\hsize}{!}{\includegraphics[clip=]{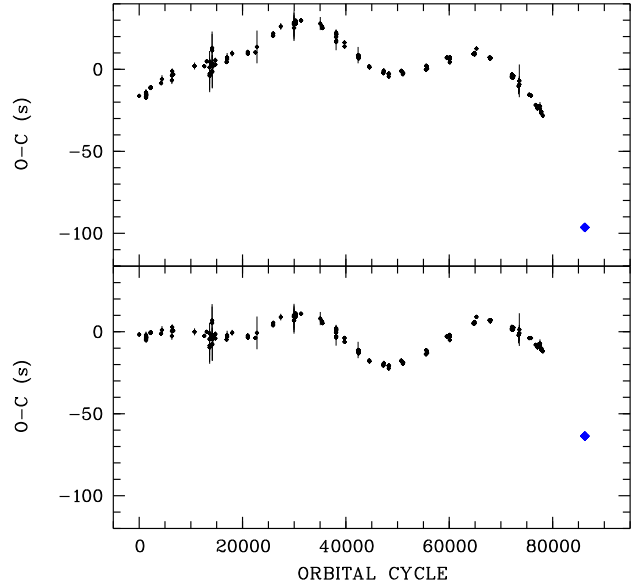}}
\caption{Observed minus calculated times of eclipse egress of \hu .
Data given in \cite{schwope+01,schwarz+09,godz+12} are shown with small 
circles, the new result obtained here is shown with a rhomb. In the upper panel 
the residuals are shown with respect to a linear ephemeris, in the lower panel 
with respect to a quadratic ephemeris.}
\label{f:omc}
\end{figure}

\section{Analysis and results}
\subsection{Eclipse parameters}
All timing data were put on the same time scale as previous data. The centers
of individual CCD frame times were corrected for arrival at the solar system barycenter
and were put on an atomic time scale by adding 32.184\,s for the conversion 
from TAI to TDT, and 35\,s, the current amount of leap seconds. 

Initially, the linear part of the two-planet LTT fit by \cite{godz+12}, 
${\rm BJD(TDB)} = 2449102.92004 + E \times 0.086820400$,
was used to convert times to binary phase.
All 7779 data points were phase-folded
and binned into 500 phase bins to generate the light curve displayed in Fig.~\ref{f:lcfol}. 
The relative brightness of the object and certain features of the light curve 
which are also labeled in the figure clearly 
indicate a high accretion state \citep[for comparison see the collection of light curves 
obtained in intermediate and high states in][their Fig.~3]{schwope+01}. The features 
are: 
(a) a pronounced pre-eclipse dip well separated from the eclipse and 
centered about 0.14 phase units before eclipse centre;
(b) a short ingress phase of the accretion stream/curtain after eclipse of the 
white dwarf; (c) the absence of a flat-bottom faint phase. Both the centre phase 
of the pre-eclipse dip, $\Delta\phi = -0.140\pm0.002$ and the phase of half intensity during 
dip ingress, $\Delta\phi = -0.175\pm0.002$ (both measured with respect to the centre of the 
eclipse), are comparable to the highest accretion state reported by \cite{schwope+01}.
Those early phases of the pre-eclipse dip indicate a late coupling of the accretion stream 
onto the magnetic field due to the high ram pressure of the accretion stream.

The times of individual ingress and egress were measured by averaging a few data 
points before and after ingress/egress, computing the half-light intensity and reading 
the times with a cursor from a graph of the light curve. Figure~\ref{f:meas} illustrates the 
procedure for data obtained on October 24.
Uncertainties of individual measurements were set to half the time resolution 
achieved, unless a data gap deteriorates the timing accuracy further.
All ingress and egress timings corrected for  the bias of 1.27\,s are listed in Table~\ref{t:ecl}.

\begin{table}[t]
\caption{Times of eclipse ingress and egress. Times are given in BJD(TDB). Errors
given do not include the systematic uncertainty of 0.5\,s.
\label{t:ecl}
}
\begin{tabular}{crlrlr}
\hline
Date & Cycle & $t_{\rm i}$ & $\Delta t_{\rm i}$ & $t_{\rm e}$& $\Delta t_{\rm e}$ \\
(MMDD) & & +2456588 & [s] & +2456588 & [s] \\
\hline
1022 & 86217 & 0.307269 &   3.0 & 0.313872 &  3.0 \\
1023 & 86229 & 1.348875 & 15.0 &                 &        \\
1024 & 86240 & 2.303966 &   3.5 & 2.310700 &  3.5 \\
1024 & 86241 & 2.390769 &   3.5 & 2.397546 &  3.5 \\
1026 & 86263 &                 &         & 4.307468 &  8.0 \\
1026 & 86264 & 4.387604 &   5.0 &                 &        \\
1030 & 86309 & 8.294531 &   3.5 & 8.301313 &  5.0 \\
1030 & 86310 & 8.381423 &   8.0 & 8.38839   &  30  \\
\hline
\end{tabular}
\end{table}

The times in the list were converted to binary cycle, the integer part was subtracted and
phases of ingress and egress were computed as weighted averages of individual measurements, 
$\left\langle \phi_{\rm i} \right\rangle  = -0.08462 \pm 0.00023$ and 
$\left\langle \phi_{\rm e}\right\rangle = -0.00715 \pm 0.00026$, respectively. 
The measured eclipse length is $581.1 \pm 3.7$\,s. This value is consistent
with the eclipse length in the 1993 high state, 584.6\,s \citep{schwope+97}, 
measured with the high-speed photometer MCCP attached to the Calar 
Alto 2.2m telescope. The MCCP-data were obtained with high time resolution that 
was sufficient to resolve eclipse ingress and egress.
 
The mean epoch of eclipse egress determined here is
BJD(TDB) $= 2456590.397536 \pm 0.000026$ for cycle 86241.
The average value for the named cycle is different by 1.7\,s from the actually 
measured value.

\subsection{Eclipse ephemeris}
The newly determined time for the eclipse egress was combined with those reported previously 
in the literature \citep{schwope+01, schwarz+09,godz+12}. The data obtained by \cite{qian+11} were not 
included because these were shown to be offset from data obtained at similar
epochs for an unknown reason \citep[see the discussion 
in][]{godz+12}. Inclusion of those data does not change our results significantly.

A weighted linear
regression to all 162 data points yields the linear ephemeris for the eclipse egress 
\begin{eqnarray*}
{\rm BJD(TDB)} =  2449102.9201891(12) + \\ E \times 0.086820403874(23)
\end{eqnarray*}
(numbers in parenthesis give formal 1$\sigma$ uncertainties, 
reduced $\chi_\nu^2=1583$ for 160 d.o.f.). The residuals with respect 
to this linear fit are shown in the upper panel of Fig.~\ref{f:omc}.
The latest data point derived in the current paper
indicates a very significant apparent decay of the orbital period
over the last two years so that the eclipse is early by $95.3\pm2.0$\,s with respect 
to the best-fit linear ephemeris and by $67.0\pm2.2$\,s with respect to the 
last data point presented by \cite{godz+12}.  When the data of \cite{qian+11} are 
included in the fit, the parameters are the same within the errors, the estimated 
errors remain unchanged, only the goodness of fit statistics is changed slightly,
reduced $\chi_\nu^2=1490$ for 170 d.o.f.. 

During the time interval 1993  -- 2011 the derivative $|{\rm d}(O-C) / {\rm d}P|$
remained below $0.7\times 10^{-7}$, the average value for the last two years was 
$1.1 \times 10^{-7}$.

A quadratic fit of the form ${\rm BJD(TDB)} = T_0 + P E + 1/2 P \dot{P} E^2$ 
reveals a better but still completely unsatisfactory representation of the data: 
$T_0 = 
2449102.9200230\pm 
            0.0000014
$, $P = 
0\fd08682042449  \pm 
0\fd00000000010
$, and 
$ \dot{P} =
(-616.1\pm 2.9) \times 10^{-14}
$
(reduced $\chi_\nu^2= 417$ for 159 d.o.f.). 
The residuals of the quadratic fit are also shown in Fig.~\ref{f:omc}.
The latest data point is early by $62.8\pm2.0$\,s with respect 
to the best-fit quadratic ephemeris and by $50.0\pm2.2$\,s with respect to the 
last data point presented by \cite{godz+12}. 
We note that the quadratic fit may be regarded just as a parameterization of our 
ignorance of the underlying behavior of the source and the 
relevant physical processes. The usual suspectives that 
could be made responsible for a quadratic term, the mechanism 
described by \cite{applegate92} and magnetic braking through 
gravitational radiation, seemingly do not provide sufficient angular momentum 
loss to explain the large observed rates \citep[see e.g.~the 
discussion in][]{vogel+08, schwarz+09}.

The new linear and quadratic ephemerides differ from the 
last published values by \cite{godz+12}. Using our formal 1$\sigma$ uncertainties
the differences appear to be real, using the uncertainties derived by \cite{godz+12}
only the value of the quadratic term appears to be significantly different. Their 
implied period derivative, $\dot{P} = 705\pm14$\,s s$^{-1}$, from their 
fitted value $\beta = 0.5 P \dot{P} = (3.06\pm0.06) \times 10^{-13}$ day cycle$^{-2}$, 
indicates an even larger period change than our quadratic fit.
However, the parameters and uncertainties derived in the two papers were
derived for different models, the model by \cite{godz+12} taking the LTT-effect of two planets 
with degenerate parameters into account. A more meaningful comparison of the 
parameters would benefit from the application of the same model.

\section{Discussion and conclusion}
The latest addition to the measured set of eclipse timings has revealed a 
very significant apparent decay of the orbital period. It is the first reported measurement 
of eclipse egress after two years without new data. We checked the reliability 
of the new data point by inspecting the data obtained quasi-simultaneously 
with XMM-Newton and found both data sets in agreement within a few seconds. 
Any possible mis-match of optical and X-ray timing data is too small 
to explain the huge trend in the $(O-C)$ data shown in both panels of Fig.~\ref{f:omc}.
The results of the satellite data and a more thorough 
discussion of the relative timing between X-ray and optical data will be 
will be published elsewhere (Schwope et al.~2014, in preparation).

The new data point lies outside the $(O-C)$ ranges of recent studies \citep{hinse+12,godz+12,wittenmyer+12}.
\cite{hinse+12} were discussing a two-planet model with a summed amplitude of
$23.4\pm0.1$\,s, too small, to make it compliant with our large $(O-C)$s. Their 
best-fit model also predicts increasingly later arrivals of the eclipse for 
cycle numbers $>$72000, contrary to what is observed.  \cite{godz+12}
were trying to fit the complete dataset with a two-planet model which gave 
apparently unphysical solutions only. Their linear model revealed massive  
third bodies, $M_3 \simeq 10$\,\msun. 
This model could probably be trimmed to reflect the large observed $(O-C)$ values, 
such a solution, however, would be mainly of academic interest. 
The quadratic model reveals an outer planet with eccentricity close to one, which 
was regarded unphysical by \cite{godz+12} themselves. A single 
planet model was found to be viable by ignoring data not obtained with 
the OPTIMA camera. This model with an amplitude of about 15\,s, however,
is discarded by our result, which also discards the underlying assumption that the mix of data 
obtained at different wavelengths makes a meaningful fit so difficult.

The planetary parameters reported in those studies therefore need revision.

The current peak-to-peak amplitudes of the $(O-C)$ curves 
are 74\,s and 125\,s for the quadratic and linear fits, respectively.  
The LTT effect of a third body in a circular orbit seen edge-on 
is 
\begin{equation} 
\Delta T \simeq \frac{2M_3G^{1/3}}{c}\left(\frac{P_3}{2\pi(M_1+M_2)}\right)^{2/3},
\end{equation}
with $M_1, M_2, M_3$ being the masses of the three bodies and $P_3$ the 
period of the third body around the binary \citep{pribulla+12}. 
Taken the measured peak-to-peak amplitudes at face value, 
planetary masses of 22 and 38 Jupiter masses for an assumed 
period of 6.9 years are derived.
Such masses are significantly larger than the typically reported 
(non-pathological) planets around 5 Jupiter masses of past studies
but should be treated with great caution; the parameters are not well constrained 
by our study. As the above equation shows, the mass scales with the 
timing amplitude which is not really measured but estimated from the current 
peak-to-peak amplitude of the timing residuals.

The whole data set seems difficult to be reconciled with a single circumbinary planet
due to the lack of regularity.
On the other hand the data are not sufficient to finally constrain the parameters 
of a putative planetary system and to test the significance, the size,  and physical 
interpretation of a quadratic term. This needs a much longer time base, 
in particular one needs to observe the turn-up towards later $(O-C)$ values.

A note on the timing errors might be in place here. The current fit results 
are dominated by the OPTIMA data points with timing uncertainties as small 
as 0.11\,s. This is smaller than the accretion area, hence at that precision a 
timing jitter due to accretion filaments or other forms of in-stationary accretion 
becomes relevant. Without a correction from the observed time of eclipse egress to 
eclipse center or without an extra error term which accounts for the finite size of the 
accretion area, searches for planets might be biased by the high 
precision of some of the data, which do not reflect the LTT effect of a 
circumbinary planet (planetary system) but the effects of in-stationary accretion. 

The bottom line of this short communication is:
The most recent addition to the $(O-C)$-data show an accelerated 
evolution of the timing residuals that rule out past models of a planet 
or a planetary system. HU Aqr deserves further monitoring with high cadence. 
\cite{godz+12} have ignored data obtained at other than optical wavelengths 
to make their one-planet model applicable. The measurement presented here 
shows that this self-restriction doesn't save a one-planet model. It appears more advantageous 
to use the complete data set for future modeling.
The current data were obtained with private equipment operated in a semi-professional 
manner. They were obtained with a comparatively small telescope 
(by professional standards) located at a mildly light-polluted site. Nevertheless
a highly constraining data point could be determined.
More advanced amateurs with similar equipment might become interested
to join the enterprise of searching for long-period planets via the LTT effect
in collaboration with professionals.
Potential targets and work in this direction are described by \cite{pribulla+12}
and \cite{backhaus+12}.

\acknowledgement
We thank our referee for constructive criticism that help to improve
the quality of the paper and Iris Traulsen and Robert Schwarz for help with the data 
reduction and useful discussions.

\bibliographystyle{aa}
\bibliography{anhutf}

\begin{thebibliography}{16}
\expandafter\ifx\csname natexlab\endcsname\relax\def\natexlab#1{#1}\fi

\bibitem[{{Applegate}(1992)}]{applegate92}
{Applegate}, J.~H. 1992, \apj, 385, 621

\bibitem[{{Backhaus} {et~al.}(2012){Backhaus}, {Bauer}, {Beuermann}, {Diese},
  {Dreizler}, {Hessman}, {Husser}, {Klapdohr}, {M{\"o}llmanns},
  {Sch{\"u}necke}, {Dette}, {Dubbert}, {Miosga}, {Rochus Vogel}, {Simons},
  {Biriuk}, {Debrah}, {Griemens}, {Hahn}, {M{\"o}ller}, {Pawlowski},
  {Schweizer}, {Speck}, {Zapros}, {Bollmann}, {Habermann}, {Haustovich},
  {Lauser}, {Liebing}, {Niederstadt}, {Hoppen}, {Kindermann}, {K{\"u}ppers},
  {Rauch}, {Althoff}, {Horstmann}, {Kellerman}, {Kietz}, {Nienaber}, {Sauer},
  {Secci}, \& {W{\"u}llner}}]{backhaus+12}
{Backhaus}, U., {Bauer}, S., {Beuermann}, K., {et~al.} 2012, A\&A, 538, A84

\bibitem[{{Go{\'z}dziewski} {et~al.}(2012){Go{\'z}dziewski}, {Nasiroglu},
  {S{\l}owikowska}, {Beuermann}, {Kanbach}, {Gauza}, {Maciejewski}, {Schwarz},
  {Schwope}, {Hinse}, {Haghighipour}, {Burwitz}, {S{\l}onina}, \&
  {Rau}}]{godz+12}
{Go{\'z}dziewski}, K., {Nasiroglu}, I., {S{\l}owikowska}, A., {et~al.} 2012,
  \mnras, 425, 930

\bibitem[{{Hakala} {et~al.}(1993){Hakala}, {Watson}, {Vilhu}, {Hassall},
  {Kellett}, {Mason}, \& {Piirola}}]{hakala+93}
{Hakala}, P.~J., {Watson}, M.~G., {Vilhu}, O., {et~al.} 1993, \mnras, 263, 61

\bibitem[{{Hinse} {et~al.}(2012){Hinse}, {Lee}, {Go{\'z}dziewski},
  {Haghighipour}, {Lee}, \& {Scullion}}]{hinse+12}
{Hinse}, T.~C., {Lee}, J.~W., {Go{\'z}dziewski}, K., {et~al.} 2012, \mnras,
  420, 3609

\bibitem[{{Horner} {et~al.}(2011){Horner}, {Marshall}, {Wittenmyer}, \&
  {Tinney}}]{horner+11}
{Horner}, J., {Marshall}, J.~P., {Wittenmyer}, R.~A., \& {Tinney}, C.~G. 2011,
  \mnras, 416, L11

\bibitem[{{Portegies Zwart}(2013)}]{port-zwart13}
{Portegies Zwart}, S. 2013, \mnras, 429, L45

\bibitem[{{Pribulla} {et~al.}(2012){Pribulla}, {Va{\v n}ko}, {Ammler-von Eiff},
  {Andreev}, {Aslant{\"u}rk}, {Awadalla}, {Balu{\v d}ansk{\'y}}, {Bonanno},
  {Bo{\v z}i{\'c}}, {Catanzaro}, {{\c C}elik}, {Christopoulou}, {Covino},
  {Cusano}, {Dimitrov}, {Dubovsk{\'y}}, {Eigmueller}, {Esmer}, {Frasca},
  {Hamb{\'a}lek}, {Hanna}, {Hanslmeier}, {Kalomeni}, {Kjurkchieva},
  {Krushevska}, {Kudzej}, {Kundra}, {Kuznyetsova}, {Lee}, {Leitzinger},
  {Maciejewski}, {Moldovan}, {Morais}, {Mugrauer}, {Neuh{\"a}user},
  {Niedzielski}, {Odert}, {Ohlert}, {{\"O}zavc{\i}}, {Papageorgiou},
  {Parimucha}, {Poddan{\'y}}, {Pop}, {Raetz}, {Raetz}, {Romanyuk}, {Ru{\v
  z}djak}, {Schulz}, {{\c S}enavc{\i}}, {Srdoc}, {Szalai}, {Sz{\'e}kely},
  {Sudar}, {Tezcan}, {T{\"o}r{\"u}n}, {Turcu}, {Vince}, \&
  {Zejda}}]{pribulla+12}
{Pribulla}, T., {Va{\v n}ko}, M., {Ammler-von Eiff}, M., {et~al.} 2012,
  Astronomische Nachrichten, 333, 754

\bibitem[{{Qian} {et~al.}(2011){Qian}, {Liu}, {Liao}, {Li}, {Zhu}, {Dai}, {He},
  {Zhao}, {Zhang}, \& {Li}}]{qian+11}
{Qian}, S.-B., {Liu}, L., {Liao}, W.-P., {et~al.} 2011, \mnras, 414, L16

\bibitem[{{Schwarz} {et~al.}(2009){Schwarz}, {Schwope}, {Vogel}, {Dhillon},
  {Marsh}, {Copperwheat}, {Littlefair}, \& {Kanbach}}]{schwarz+09}
{Schwarz}, R., {Schwope}, A.~D., {Vogel}, J., {et~al.} 2009, A\&A, 496, 833

\bibitem[{{Schwope} {et~al.}(2011){Schwope}, {Horne}, {Steeghs}, \&
  {Still}}]{schwope+11}
{Schwope}, A.~D., {Horne}, K., {Steeghs}, D., \& {Still}, M. 2011, A\&A, 531,
  A34

\bibitem[{{Schwope} {et~al.}(1997){Schwope}, {Mantel}, \& {Horne}}]{schwope+97}
{Schwope}, A.~D., {Mantel}, K., \& {Horne}, K. 1997, A\&A, 319, 894

\bibitem[{{Schwope} {et~al.}(2001){Schwope}, {Schwarz}, {Sirk}, \&
  {Howell}}]{schwope+01}
{Schwope}, A.~D., {Schwarz}, R., {Sirk}, M., \& {Howell}, S.~B. 2001, A\&A,
  375, 419

\bibitem[{{Schwope} {et~al.}(1993){Schwope}, {Thomas}, \&
  {Beuermann}}]{schwope+93}
{Schwope}, A.~D., {Thomas}, H.~C., \& {Beuermann}, K. 1993, A\&A, 271, L25

\bibitem[{{Vogel} {et~al.}(2008){Vogel}, {Schwope}, {Schwarz}, {Kanbach},
  {Dhillon}, \& {Marsh}}]{vogel+08}
{Vogel}, J., {Schwope}, A., {Schwarz}, R., {et~al.} 2008, in American Institute
  of Physics Conference Series, Vol. 984, High Time Resolution Astrophysics:
  The Universe at Sub-Second Timescales, ed. D.~{Phelan}, O.~{Ryan}, \&
  A.~{Shearer}, 264--267

\bibitem[{{Wittenmyer} {et~al.}(2012){Wittenmyer}, {Horner}, {Marshall},
  {Butters}, \& {Tinney}}]{wittenmyer+12}
{Wittenmyer}, R.~A., {Horner}, J., {Marshall}, J.~P., {Butters}, O.~W., \&
  {Tinney}, C.~G. 2012, \mnras, 419, 3258

\end{thebibliography}
  
\end{document}